\documentstyle[psfig,times]{mn}
\newread\epsffilein    
\newif\ifepsffileok    
\newif\ifepsfbbfound   
\newif\ifepsfverbose   
\newdimen\epsfxsize    
\newdimen\epsfysize    
\newdimen\epsftsize    
\newdimen\epsfrsize    
\newdimen\epsftmp      
\newdimen\pspoints     
\def\epsfbox#1{\global\def\epsfllx{72}\global\def\epsflly{72}%
   \global\def\epsfurx{540}\global\def\epsfury{720}%
   \def\lbracket{[}\def\testit{#1}\ifx\testit\lbracket
   \let\next=\epsfgetlitbb\else\let\next=\epsfnormal\fi\next{#1}}%
\def\epsfgetlitbb#1#2 #3 #4 #5]#6{\epsfgrab #2 #3 #4 #5 .\\%
   \epsfsetgraph{#6}}%
\def\epsfnormal#1{\epsfgetbb{#1}\epsfsetgraph{#1}}%
\def\epsfgetbb#1{%
\openin\epsffilein=#1
\ifeof\epsffilein\errmessage{I couldn't open #1, will ignore it}\else
   {\epsffileoktrue \chardef\other=12
    \def\do##1{\catcode`##1=\other}\dospecials \catcode`\ =10
    \loop
       \read\epsffilein to \epsffileline
       \ifeof\epsffilein\epsffileokfalse\else
          \expandafter\epsfaux\epsffileline:. \\%
       \fi
   \ifepsffileok\repeat
   \ifepsfbbfound\else
    \ifepsfverbose\message{No bounding box comment in #1; using defaults}\fi\fi
   }\closein\epsffilein\fi}%
\def\epsfsetgraph#1{%
   \epsfrsize=\epsfury\pspoints
   \advance\epsfrsize by-\epsflly\pspoints
   \epsftsize=\epsfurx\pspoints
   \advance\epsftsize by-\epsfllx\pspoints
   \epsfxsize\epsfsize\epsftsize\epsfrsize
   \ifnum\epsfxsize=0 \ifnum\epsfysize=0
      \epsfxsize=\epsftsize \epsfysize=\epsfrsize
     \else\epsftmp=\epsftsize \divide\epsftmp\epsfrsize
       \epsfxsize=\epsfysize \multiply\epsfxsize\epsftmp
       \multiply\epsftmp\epsfrsize \advance\epsftsize-\epsftmp
       \epsftmp=\epsfysize
       \loop \advance\epsftsize\epsftsize \divide\epsftmp 2
       \ifnum\epsftmp>0
          \ifnum\epsftsize<\epsfrsize\else
             \advance\epsftsize-\epsfrsize \advance\epsfxsize\epsftmp \fi
       \repeat
     \fi
   \else\epsftmp=\epsfrsize \divide\epsftmp\epsftsize
     \epsfysize=\epsfxsize \multiply\epsfysize\epsftmp   
     \multiply\epsftmp\epsftsize \advance\epsfrsize-\epsftmp
     \epsftmp=\epsfxsize
     \loop \advance\epsfrsize\epsfrsize \divide\epsftmp 2
     \ifnum\epsftmp>0
        \ifnum\epsfrsize<\epsftsize\else
           \advance\epsfrsize-\epsftsize \advance\epsfysize\epsftmp \fi
     \repeat     
   \fi
   \ifepsfverbose\message{#1: width=\the\epsfxsize, height=\the\epsfysize}\fi
   \epsftmp=10\epsfxsize \divide\epsftmp\pspoints
   \newcount\figskipcount
      \message{#1 \the\epsfysize  }
   \vbox to\epsfysize{\vfil\hbox to\epsfxsize{%
      \includegraphics{#1}%
      \hfil}}%
\epsfxsize=0pt\epsfysize=0pt}%

{\catcode`\%=12 \global\let\epsfpercent=
\long\def\epsfaux#1#2:#3\\{\ifx#1\epsfpercent
   \def\testit{#2}\ifx\testit\epsfbblit
      \epsfgrab #3 . . . \\%
      \epsffileokfalse
      \global\epsfbbfoundtrue
   \fi\else\ifx#1\par\else\epsffileokfalse\fi\fi}%
\def\epsfgrab #1 #2 #3 #4 #5\\{%
   \global\def\epsfllx{#1}\ifx\epsfllx\empty
      \epsfgrab #2 #3 #4 #5 .\\\else
   \global\def\epsflly{#2}%
   \global\def\epsfurx{#3}\global\def\epsfury{#4}\fi}%
\def\epsfsize#1#2{\epsfxsize}
\begin{document}
\title[
The AGN contribution to deep submillimetre surveys and the far-infrared
background
]
{
The AGN contribution to deep submillimetre surveys and the far-infrared
background
}
\author[O. Almaini,  A. Lawrence, B.J. Boyle]
{O.~Almaini,$^{1}$
A. Lawrence,$^1$ 
B.J. Boyle$^2$ \\
$^1$Institute for Astronomy, 
University of Edinburgh, 
Royal Observatory, 
Blackford Hill, Edinburgh EH9 3HJ \\
$^2$Anglo-Australian Observatory, PO Box 296, Epping, NSW2121, Australia}

\date{Accepted 1999 March 24. Received 1999 March 15; in original form
1998 September 2} \maketitle

\begin{abstract}
A great deal of interest has been generated recently by the results of
deep submillimetre surveys, which in principle allow an unobscured
view of dust-enshrouded star formation at high redshift.  The
extragalactic far-infrared and submillimetre backgrounds have also
been detected, providing further constraints on the history of star
formation.  In this paper we estimate the fraction of these
backgrounds and source counts that could be explained by AGN.  The
relative fractions of obscured and unobscured objects are constrained
by the requirement that they fit the spectrum of the cosmic X-ray
background.  On the assumption that the spectral energy distributions
of high redshift AGN are similar to those observed locally, we find
that one can explain $10-20$  per cent of the $850\mu$m SCUBA sources at
$1$mJy and a similar fraction of the far-infrared/submillimetre
background. The exact contribution depends on the assumed cosmology
and the space density of AGN at high redshift ($z>3$), but we conclude
that active nuclei will be present in a significant (though not
dominant) fraction of the faint SCUBA sources. This fraction could be
significantly higher if a large population of AGN are highly obscured
(Compton-thick) at X-ray wavelengths.

\end{abstract}


\section{Introduction}

Deep submillimetre observations offer the potential to revolutionise
our understanding of the high redshift Universe.  Longward of
100$\mu$m, both starburst galaxies and AGN show a very steep decline
in their continuum emission, which leads to a large negative
K-correction as objects are observed with increasing redshift. This
effectively overcomes the `inverse square law' to pick out the most
luminous objects in the Universe to very high redshift (Blain \&
Longair 1993).  Since the commissioning of the SCUBA array at the
James Clerk Maxwell Telescope a number of groups have announced the
results from deep submillimetre surveys, all of which find a high
surface density of sources at $850 \mu$m (Smail et al. 1997, Hughes et
al 1998, Barger et al. 1998, Eales et al. 1998, Blain et al. 1999b).  The
implication is the existence of a large population of hitherto
undetected dust enshrouded galaxies.  In particular, the implied
star-formation rate at high redshift ($z>2$) could be significantly
higher than that deduced from uncorrected optical-UV observations
(Hughes et al. 1998).  The recent detections of the
far-infrared/submillimetre background by the DIRBE and FIRAS
experiments (Puget et al. 1996; Fixsen et al. 1998; Hauser et al. 1998)
provide further constraints, representing the integrated far-infrared
emission over the entire history of the Universe (Dwek et al. 1998).
Since most of this background has now been resolved into discrete
sources by SCUBA, the implication is that most high
redshift star-forming activity occurred in rare, exceptionally
luminous systems.

In this paper we investigate the possible contribution from AGN to
both the far-infrared/submillimetre background and the SCUBA source
counts at $850\mu$m.  The deep SCUBA sources in particular are
believed to be the high redshift equivalents to local Ultra-Luminous
Infrared Galaxies (ULIRGs; Sanders \& Mirabel 1996) and hence a large
AGN fraction may not be a surprise. Our strategy is to predict faint
submillimetre counts based on our knowledge of the AGN luminosity
function and its evolution, together with an assumed spectral energy
distribution (SED). We take account of the likely population of {\em
obscured} AGN, using models which reproduce the spectrum of the hard
X-ray background (XRB).

\section{Obscured AGN and the X-ray background}

Locally it is clear that a large fraction of AGN are obscured by gas
and dust. Obscured (e.g.  narrow-line) AGN are believed to outnumber
unobscured (e.g. broad-line) objects by a factor of $\sim$ a few, with
significant uncertainties depending on the selection techniques and
the assumptions made (Lawrence 1991; Huchra \& Burg 1992; Osterbrock
\& Martel 1993).  It is now important to understand whether this ratio
continues to high redshifts and luminosities.  Radio emission is
unaffected by obscuration, and among radio-loud objects at least there
is clear evidence that narrow-line radio galaxies are common to very
high powers and redshifts.  Furthermore, recent deep X-ray surveys
using ROSAT, ASCA and Beppo-SAX have detected substantial numbers of
`narrow-line X-ray galaxies', many of which show very clear evidence
for obscured AGN activity (Boyle et al. 1995; Almaini et al. 1995;
Ohta et al. 1996; Iwasawa et al. 1997; Boyle et al. 1998; Schmidt et
al. 1998, Fiore et al. 1999). It seems likely that this is the `tip of
the iceberg' of a large population of obscured AGN, full confirmation
of which will soon be possible with the next generation of X-ray
satellites (e.g. AXAF and XMM).

There is increasing evidence that these obscured AGN are responsible
for the production of the hard XRB.  At soft X-ray energies (below
2keV) almost all of the XRB has been resolved into discrete sources,
most of which turn out to be broad-line QSOs (Shanks et al. 1991), but
the X-ray spectra of ordinary QSOs are too steep to explain the XRB at
higher energies. The energy density of the XRB actually peaks at $\sim
30$keV, where ordinary broad-line QSOs can account for only $\sim 20$
per cent (Fabian et al. 1998).  Obscured AGN provide a very natural
explanation, since photoelectric absorption allows only the hard
X-rays to penetrate (Setti \& Woltjer 1989). Models have been
developed which provide very good fits to  the XRB spectrum, X-ray
source counts and the local distribution of absorbing columns
(Comastri et al. 1995; Miyaji et al. 1998).  The implication is that
most of the energy density generated by accretion in the Universe
takes place in obscured AGN.  As outlined by Fabian \& Iwasawa (1999)
this hidden population could explain the apparent discrepancy between
predicted present day black hole densities and the observations of
Magorrian et al. (1998).

\section{The AGN luminosity function}

\subsection{Type 1 AGN}

Our predictions are based on the X-ray Luminosity Function (XLF) of
Boyle et al (1994), which we take to represent the baseline population
of unobscured (Type 1) AGN.  Since the optical/UV spectra of these AGN
all show broad emission lines, and the X-ray spectra show no evidence
for photoelectric absorption (Almaini et al. 1996) we can safely
assume that this sample includes only relatively unobscured objects
(e.g. $N_H< 10^{22}$cm$^{-1}$).  This luminosity function was modelled
with a broken power law of the form:

\begin{equation}
\Phi_X(L_X)= \left\{\begin{array}{ll}
\Phi_X^*L_{44}^{-\gamma_1} & L_X<L_X^* \\
& \\
\frac{\Phi_X^*}{L_{44}^{*(\gamma_1-\gamma_2)}}L_{44}^{-\gamma_2}
& L_X>L_X^* \\
\end{array}
\right.
\end{equation}

\noindent where $\gamma_1$ and $\gamma_2$ represent the faint and
bright end slopes of the XLF respectively and $L_{44}$ is the
$0.3-3.5\,$keV luminosity expressed in units of $10^{44}$erg
s$^{-1}$. A good fit to the evolution of this luminosity function was
obtained with a Pure Luminosity Evolution (PLE) model out to a maximum
redshift $z_{max}$:

\begin{equation}
L_X^*(z) = \left\{\begin{array}{ll}
 L_X^*(0) (1+z)^k & z<z_{max}  \\
& \\
L_X^*(z_{max}) & z>z_{max}  
\end{array}
\right.
\end{equation}

The best fitting parameters were $z_{max}=1.6(1.79)$, $k=3.25(3.34)$,
$\gamma_1=1.36(1.53)$, $\gamma_2=3.37(3.38)$,
$L_X^*=10^{43.57(43.70)}$, and $\Phi_X^*=1.59(0.63) \times 10^{-6}$
Mpc$^{-3}$ for a cosmology with
$q_0=0.5(0.0)$ and $H_0=50\,$km s$^{-1}$Mpc$^{-1}$.  Further details
can be found in Boyle et al. (1994) (models S \& T).

\subsection{Correction for obscured AGN}

To constrain the relative fractions of obscured and unobscured objects
we use the Comastri et al. (1995) population synthesis model for the
XRB. In this model a population of obscured AGN with a range of column
densities is added to the unobscured population in order to obtain
good fits to the XRB spectrum and the X-ray source count
distribution. Excellent fits to the hard XRB were obtained by assuming
a simple model in which the space density of obscured objects and the
distribution in obscuring column densities are free parameters.  The
best fit was obtained with a ratio of obscured to unobscured AGN ($N_H
< 10^{22}$cm$^{-1}$) in the range $2.4-3.7$, in very good agreement
with local observations (see Section 2).  Although the affects of the
photoelectric absorption on the X-ray source counts are complicated,
in the submillimetre regime the gas and dust is transparent. We
therefore simply scale the unobscured QSO luminosity function as
follows:

\begin{equation}
\Phi_{obscured} = 3 \times \Phi_{unobscured}
\end{equation}

\subsection{The high redshift evolution}

At the highest redshifts ($z>z_{max}$) the X-ray luminosity function
of Boyle et al. (1994) is formally consistent with being constant, but
beyond $z=3$ there are very few QSOs in this survey.  Optical surveys
for bright QSOs find evidence for an exponential $\em decline$ in the
QSO space density towards high $z$, parameterised as follows by
Schmidt, Schneider \& Gunn (1995):

\begin{equation}
\Phi(z) = \Phi(2.7)e^{(2.7-z)} \hspace{1.2cm} z>2.7
\end{equation}

Similar results were found by Hewett et al. (1993).  The reality of
this exponential decline is unclear, however.  We note that these
optically derived surveys may be seriously underestimating the high
redshift QSO population because of intervening absorption along the
line of sight. This is discussed further in Section 5.  We therefore
consider two models for the high redshift evolution.  For Model A we
use the exponential decline given by Equation 4, while for Model B we
consider the possibility that the space density of QSOs is constant to
$z=5$ (with an exponential decline thereafter), in good agreement with
the findings of the Ultra-Deep X-ray Survey by Hasinger  (1998).

\section{The thermal far-infrared spectral template}

In radio-quiet AGN there is significant evidence that the
far-infrared/submillimetre emission is dominated by thermal
re-radiation from dust (Carleton et al. 1987, Hughes et al. 1993).  To
model this spectrum, we use the best fitting spectral energy
distribution obtained by Hughes, Davies \& Ward (1999), who have
obtained the largest collection of far-infrared and submillimetre
observations of local AGN.  They find that the $50-1300\mu$m emission
is consistent with thermal re-radiation from dust at a temperature of
$35-40$K. This can be modelled by an isothermal grey-body curve:

\begin{equation}
f\nu \propto \frac{\nu^{3+\beta}}{\exp(h\nu/kT) -1}  \hspace{1.2cm} \lambda > 50\mu m
\end{equation}

in which $T$ is the temperature of the dust and $\beta$ is the
emissivity index, derived by assuming that the grey-body is
transparent to its own emission.  Best fitting values were found to be
$\beta=1.7\pm0.3$ and $T=37\pm5$K (see Hughes, Davies \& Ward 1999 for
full details). We note that strikingly similar temperatures have
recently been reported for high redshift AGN by Benford et al. (1998).
At mid-infrared wavelengths they approximate the spectrum with a power
law of the form $f\nu \propto \nu^{-1.3}$.

In order to normalise these SEDs to our X-ray 
luminosity function, we use the mean far-infrared to X-ray
relationship determined for a large sample of broad-line, radio-quiet
quasars by Green, Anderson \& Ward (1992). Similar ratios were found
by Carleton et al. (1987).

\begin{equation}
f_{100\mu m} = 3.2\times10^5 f_{1keV}
\end{equation}

We note however that locally there is a  large
scatter in this X-ray to far-infrared ratio.  To allow for this we adopt a
Gaussian distribution in the log of the flux ratio given above with a
dispersion $\sigma=0.19$, as observed by Green, Anderson \& Ward (1992).

A major source of uncertainty in our method is clearly the
extrapolation of these local SEDs to AGN at high redshift.  Accurate
far-infrared and submillimetre photometry of high redshift, X-ray
selected QSOs would help to resolve this issue, which will certainly
be possible with SCUBA and future submillimetre
detectors. If these AGN contain more isotropically distributed dust
they may well turn out to be brighter at these wavelengths, which 
could boost the source predictions significantly.

\begin{figure*} 
\centerline{\psfig{figure=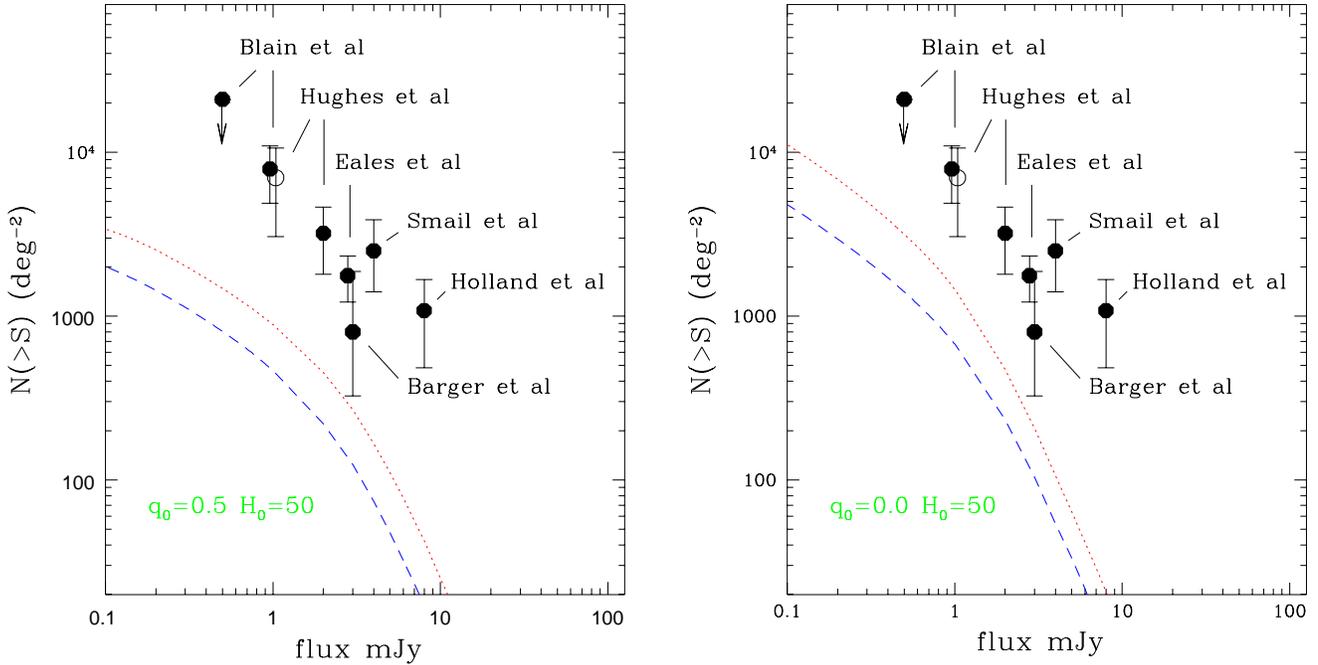,width=1.15\textwidth,angle=270}}
\caption{Predicted AGN number counts compared with recent SCUBA
measurements of the total submillimetre source density at $850\mu$m.
The unfilled circle is taken from an estimate of the source density at
1mJy based on a fluctuation analysis (Hughes et al. 1998).  In Model A
(dashed lines) we adopt the conservative assumption that the AGN space
density, $\Phi$, declines exponentially beyond $z=2.7 $. Model B
(dotted lines) assumes a constant $\Phi$ out to $z=5$.}
\end{figure*}

\section{The submillimetre count predictions}

Using the SED defined above, combined with the evolving X-ray
luminosity function from Section 3, one can predict the AGN
submillimetre counts at $850\mu$m.  The results are shown in Figure 1
for two different cosmologies, where we compare with the total SCUBA
detections from recent deep surveys.  We note that a more recent model
for the XRB has recently been produced by Miyaji et al. (1998), based
on the Ultra-Deep Lockman Hole survey of Hasinger (1998). In this
survey they find that a Luminosity Dependent Density Evolution (LDDE)
model provides a better fit to the evolution of the AGN luminosity
function then a PLE model. Using model parameters provided by Miyaji
(private communication) we find submillimetre counts predictions which
are approximately $40$ per cent higher than shown in Figure 1 at the
flux densities of the SCUBA surveys.

In using these luminosity functions, we integrate over the luminosity
range $10^{42} < L_X < 10^{48}$ erg s$^{-1}$. Integrating to higher
luminosities make no significant difference, due to the very low space
density of such sources.  There is, of course, significant evidence
that AGN activity continues to fainter luminosities (e.g. Ho,
Filippenko \& Sargent 1997) but this also makes no difference to our
predictions. At an $850\mu$m limit of 1mJy there is no significant
contribution from QSOs below a luminosity of $L_x \sim 10^{43}$ erg
s$^{-1}$.

A major source of uncertainty in determining these predictions is the
space density of AGN at high redshift ($z>3$). For Model $A$ we adopt
the conservative assumption that $\Phi$ declines exponentially above
redshift $z=2.7$, in the same manner as determined for bright optical
QSOs (Schmidt, Schneider \& Gunn 1995).  To illustrate the importance
of the high redshift space density, we also plot the extreme case
where $\Phi$ does not decline beyond $z=2.7$, but instead remains
constant out to a maximum redshift of $z=5$ (Model B).  The exact
behaviour in the QSO space density at high redshift makes little
difference to the XRB but due to the negative K-correction the
predictions in the submillimetre waveband are altered considerably.
As discussed in Fall \& Pei (1993), because of dust along the line of
sight in damped Ly$\alpha$ absorption systems, existing optical QSO
surveys may be seriously incomplete at high redshift, with possibly up
to $90$ per cent  missing at $z=4$.  Radio emission is unaffected by such
absorption, and the high redshift decline seen in the radio-loud
population would seem to argue against any serious incompleteness
(Shaver et al. 1996, Dunlop 1997). We note, however, that recent
results from the deepest X-ray surveys seem to suggest
that the space density of QSOs remains $\em constant$ out to $z\sim5$
(Hasinger 1998).  We conclude that the exact form of high redshift
evolution remains uncertain, but the reality is likely to lie
somewhere between Models A and B.  It is worth noting that the $n(z)$
distribution of submillimetre selected quasars will in principle allow
a very sensitive discriminant between these competing models (see also
Blain et al. 1999a).

It is clear from Figure 1 that a sizeable fraction of the existing
submillimetre sources can be explained by AGN.  Taking the $1$mJy
detection by Blain et al. (1999b), for example, the $q_0=0.0$ Models A
and B account for 9 and 19 per cent of the sources respectively.  The
$q_0=0.5$ models give very similar source counts at brighter fluxes
but flatten considerably at the faint end, accounting for only 7 and
13 per cent of the counts at $1$mJy. The uncertainties in our
method could account for a factor of $\sim$ a few in either direction,
but nevertheless we conclude that a significant (though not dominant)
fraction of the recent SCUBA detections will contain an AGN. As
discussed in Section 7, however, these predictions are almost
certainly lower limits since they do not include Compton-thick AGN.

\section{AGN and the far-infrared/submillimetre background}

As discussed in Section 2, there is now growing evidence that AGN
could account for the majority of the X-ray background.  Recent
analysis of COBE {\it FIRAS} data has also led to the first detections
of an extragalactic component to the far-infrared/submillimetre
background ($100-1000\mu$m; Puget et al. 1996, Fixsen et al. 1998). At
present there are considerable uncertainties involved in the
subtraction of galactic emission, zodiacal light and the cosmological
microwave background, but by a variety of independent techniques
Fixsen et al. (1998) find a mean spectrum for the far-infrared
background which can be parameterised by the following function:

\begin{equation}
I_{\nu} = 1.3\pm0.4 \times10^{-5}(\nu/\nu_o)^{0.64\pm0.12}P_{\nu}(18.5\pm1.2 K)
\end{equation}

where $\nu_o=100$cm$^{-1}$ and $P_{\nu}$ is the Planck function. The
nature of this radiation is still unclear, but at $850\mu$m most of
the background has now been resolved into discrete sources by SCUBA
(e.g. see Blain et al. 1999b).  Using the models outlined above we can
explicitly calculate the potential contribution from galaxies
containing bright AGN. Extrapolating down to an $850\mu m$ flux limit
of $0.01$mJy the resulting background contributions are shown in
Figure 2. With a conservative estimate for the space density of AGN,
one can explain  $\sim 10$ per cent of the
far-infrared/submillimetre background.  With a higher space density of
AGN (Model B) the contribution increases, particularly towards longer
wavelengths, with $\sim 20$ per cent of the background at
$850\mu$m.

\section{Compton-thick AGN}

In the arguments above, we constrain the relative contributions of
obscured and unobscured AGN using population synthesis models which
reproduce the spectrum of the XRB.  Above a critical column density of
$\sim 10^{24-25}$ atom cm$^{-2}$, however, the obscuring medium can
become optically thick to X-rays by Compton scattering, leading to
almost total absorption.  Recent BeppoSAX hard X-ray observations of a
well defined sample of Seyfert 2 galaxies have revealed that the
absorbing columns are in general very much higher than previously
thought, and that $\em most$ appear to be Compton-thick (Maiolino et
al 1998). The contribution of such objects to the XRB will be
negligible, and hence the space density of AGN required to fit the
X-ray background will always be a lower limit.  Hence the predictions
shown in Figures 1 \& 2 could be substantially higher, by $\sim 50$
per cent if we adopt the distribution found by Maiolino et al.
(1998). We note that Compton-thick AGN appear to be very common among
local AGN-dominated ULIRGs (Sanders \& Mirabel 1996) which show very
high central gas densities.

\begin{figure*} 
\centerline{\psfig{figure=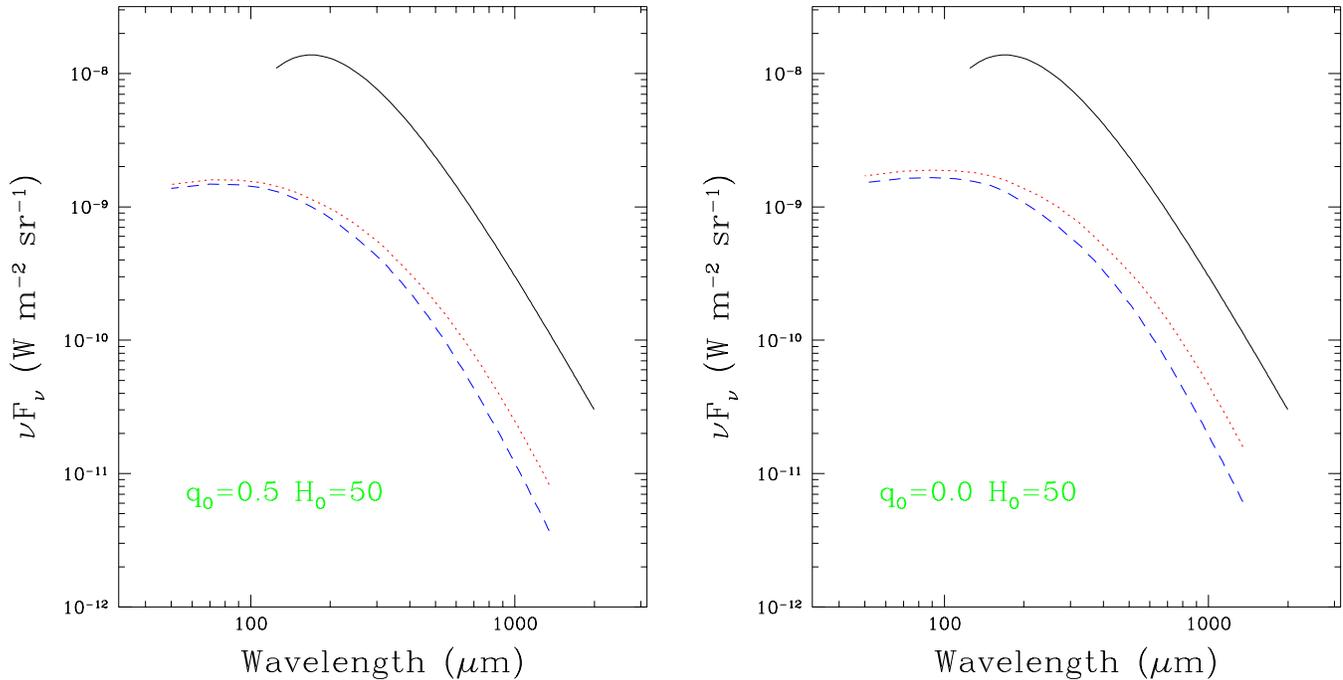,width=1.17\textwidth,angle=270}}
\caption{Predicted AGN contribution to the far-infrared/submillimetre
background.  The dashed and dotted lines are based on the Models A and
B respectively.  The solid black line represents an analytical fit to
the background spectrum obtained by Fixsen et al. (1998).}
\end{figure*}

\section{Conclusions}

There is growing evidence that obscured AGN are responsible for most
of the cosmic XRB. Using models for the XRB to constrain the relative
fractions of obscured and unobscured AGN, combined with recent
measurements of their far-infrared and submillimetre SEDs, we estimate
the contribution of AGN to the far-infrared and submillimetre
background. We also explicitly calculate the predicted source counts
at $850\mu m$, where a lot of excitement has been generated recently
from the results of deep SCUBA surveys.

Several lines of argument lead to the conclusion that a significant
fraction of the SCUBA detections will contain bright AGN.  The very
high far-infrared luminosities implied for these sources suggest that
they are high redshift equivalents to local ULIRGs. At the
luminosities of the recently discovered SCUBA sources ($L\sim
10^{12}L_{\odot}$) we note that $20-30$ per cent of local ULIRGs
show clear evidence for an AGN (e.g. Seyfert 1 or Seyfert 2), with a
further similar fraction showing more ambiguous LINER activity
(Sanders \& Mirabel 1996).  Our explicit calculations suggest that the
AGN contribution in recent SCUBA surveys is likely to be at least
$10-20$ per cent, with the exact contribution depending on the
assumed cosmology and the space density of AGN at high redshift. This
estimate is constrained by the requirement that we fit the spectrum of
X-ray background, and hence the fraction could be much higher if a
large population of AGN are Compton-thick, as recently observed by
BeppoSAX for local Seyferts (Maiolino et al. 1998). Such AGN will may
not be revealed even by AXAF or XMM.

If the dust in these AGN is heated by accretion processes, a
significant AGN contribution may help to explain the apparent
contradiction implied by the recent SCUBA surveys, namely that such
massive star-formation at early epochs could very easily over-predict
the abundance and metallicity of low-mass stars in the local Universe
(Blain et al. 1999a).  In many other ways a large AGN fraction is not a
surprise.  Of the very few high redshift submillimetre sources with
good optical spectra, at least 3 show clear evidence for the presence
of an AGN (e.g.  Ivison et al. 1998, Irwin et al. 1998, Serjeant et al.
1998).

The most puzzling question is the origin of the far-infrared emission
in AGN themselves, even if they do explain a large fraction of the
SCUBA sources. At these wavelengths the relative roles of AGN and
starburst activity remain controversial, both in IR selected ULIRGs
(Sanders \& Mirabel 1996) and in quasars (Lawrence 1997).  Is
the dust heated by the AGN or a nuclear starburst?  Both energy
sources are clearly present in a large fraction of these objects.  At
mid-infrared wavelengths the AGN probably dominates but at $100\mu$m
the situation remains unclear (Sander \& Mirabel 1996).  Hence the
emission seen in the far-infrared/submillimetre background, and from
the recently detected SCUBA sources, could be entirely due to star
forming activity, albeit with the interesting implication that a large
fraction of the star formation in the early Universe took place in the
cores of galaxies containing active quasars.

\section*{ACKNOWLEDGMENTS}

We thank Jim Dunlop, Dave Hughes, Andy Fabian and John Peacock for
useful discussions.  We are also indebted to Guenther Hasinger and
Takamitsu Miyaji for their comments and for providing us with details
of their luminosity function prior to publication.

\end{document}